\documentclass[prl,twocolumn,aps,showpacs]{revtex4}

\usepackage{graphicx}

\begin{document}

\title{Topological Aspect and Quantum Magnetoresistance of
  $\beta$-Ag$_2$Te}

\author{Wei Zhang, Rui Yu, Wanxiang Feng, Yugui Yao, Hongming Weng, Xi
  Dai, Zhong Fang}

\affiliation{Beijing National Laboratory for Condensed Matter Physics,
  and Institute of Physics, Chinese Academy of Sciences, Beijing
  100190, China;}

\date{\today}

\begin{abstract}
  To explain the unusual non-saturating linear magnetoresistance
  observed in silver chalcogenides, the quantum scenario has been
  proposed based on the assumption of gapless linear energy spectrum.
  Here we show, by first principles calculations, that
  $\beta$-Ag$_2$Te with distorted anti-fluorite structure is in fact a
  topological insulator with gapless Dirac-type surface states. The
  characteristic feature of this new binary topological insulator is
  the highly anisotropic Dirac cone, in contrast to known examples,
  such as Bi$_2$Te$_3$ and Bi$_2$Se$_3$. The Fermi velocity varies
  an order of magnitude by rotating the crystal axis.
\end{abstract}

\pacs{71.20.-b, 73.20.r, 73.43.Qt}
\maketitle

Ag$_2$Te, one of silver chalcogenides, is known as Hessite mineral in
nature. It was used as ionic conductor at high temperature $\alpha$
phase, and it undergoes a phase transition below 417 K into the
$\beta$ phase, a narrow gap semiconductor, where the ion migration is
frozen and the compound is non-magnetic. The gap of $\beta$-Ag$_2$Te
is in the range of several tens meV~\cite{gap}, the mobility of
carriers is high and the effective mass is of the order of
10$^{-2}$m$_0$ (m$_0$ is the free electron mass)~\cite{mass-1}.
Unusually large and non-saturating linear magnetoresistance (MR) were
observed in $\beta$-Ag$_{2+\delta}$Te for the field range
10$\sim$55000 Oe and temperature range 4.5$\sim$300 K~\cite{Xu}, in
contrast to the conventional theory of metal with closed Fermi
surface, from which the quadratic (low-field) and saturating
(high-field) MR is expected. This leads to the proposal of quantum MR
by Abrikosov~\cite{Abrikosov}, where only the lowest Landau level
remains occupied. However, for a quadratic energy spectrum, the Landau
level spacing (which depends on the field linearly) is about 0.1 K
estimated for $\beta$-Ag$_2$Te at 10 Oe), which is too
small~\cite{Abrikosov}. It is therefore necessary to assume the
gapless linear energy spectrum~\cite{Abrikosov}, such that the field
dependence of Landau level spacing follows the square root rule
($\Delta E_n\propto\sqrt{B}$), similar to case of Graphene.  The
linear energy spectrum may come from the strong disorder as pursued by
Abrikosov, however, we will show in this paper that $\beta$-Ag$_2$Te
is in fact a topological insulator with gapless linear Dirac-type
surface states. This raises the possibility that the observed unusual
MR may largely come from the surface/interface contributions.

Topological insulator (TI), characterized by the Z$_2$-invariance and
protected by the time-reversal symmetry, is a new state of quantum
matter~\cite{Kane_PRL_2005,Bernevig_PRL_2006,Moore,Fu1,Fu2}. It is
different with trivial insulator in the sense that its bulk is
insulating, while its surface supports metallic Dirac fermions. Exotic
quantum phenomena, such as Majorana fermion~\cite{PRL100},
magnetoelectric effect~\cite{PRB78}, and quantum anomalous Hall
effect~\cite{QAH}, are expected from TIs. It has been demonstrated
that TI can be realized in 2D systems, such as HgTe/CdTe quantum
well~\cite{Bernevig}, or 3D materials like
Bi$_{1-x}$Sb$_x$~\cite{Hsieh}. The discovery of Bi$_2$Se$_3$ family
TIs with single Dirac cone on the surface is a significant
progress~\cite{Zhang}, where the bulk gap is as large as 0.3 eV, which
makes the room temperature applications possible. There are several
more recent proposals, such as TlBiX$_2$ (X=Te, Se)
compounds~\cite{Binghai}, ternary Heusler alloys~\cite{Heusler-1},
chalcopyrites~\cite{Pyrite-1}, which all involve three or more
elements, and may require additional distortions to open up the bulk
gap. For all known TIs up to now, the surface Dirac cones are almost
isotropic and Fermi velocity is nearly a constant. Here we will shown
that highly anisotropic surface Dirac cone can be obtained in
$\beta$-Ag$_2$Te, a new binary TI with distorted anti-fluorite
structure.

\begin{figure}[tbp]
\includegraphics[clip,scale=0.4]{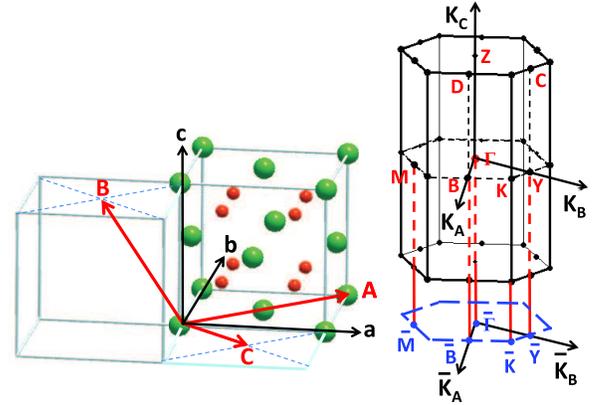}
\caption{(Color online) (a) The cubic anti-fluorite structure of
  $\alpha$-Ag$_2$Te and its structural relationship to the $\beta$
  phase. The translational vectors of $\alpha$ and $\beta$ phases are
  labeled as $a, b, c$ and $A, B, C$, respectively. (b) The Brillouin
  zone of $\beta$-Ag$_2$Te, and its projected surface BZ to the plane
  perpendicular to $C$-aixs.}
\label{structure}
\end{figure}

We calculate the electronic structures of Ag$_2$Te by using the WIEN2k
package. The generalized gradient approximation (GGA) is used for the
exchange-correlation functional, and Brillouin zone (BZ) is sampled
with 21$\times$21$\times$21 grids for $\alpha$ phase and
10$\times$10$\times$10$\times$ for $\beta$ phase. We construct the
projected atomic Wannier (PAW) functions~\cite{Wei} for $s$ and $p$
orbitals of Ag and Te. With this set of PAW bases, an effective model
Hamiltonian for a slab of 45 layers along $C$-axis is established and
the topologically nontrivial surface state is obtained from it.

The high temperature $\alpha$-phase of Ag$_2$Te can be regarded as the
anti-fluorite structure (shown in Fig. 1) in
average~\cite{Hasegawa}. Defining the cubic translational vectors as
$a, b, c$ ($|a|$=$|b|$=$|c|$), this structure can be constructed from
three fcc sub-lattices called as Te, Ag(1) and Ag(2) sub-lattice,
respectively. The Ag(1) sub-lattice is shifted from the Te sub-lattice
by a vector ($a$/4, $b$/4, $c$/4). If we only consider the Te and
Ag(1) sub-lattices, it gives the same structure as zinc blende (like
HgTe or CdTe). By adding the additional Ag(2) sub-lattice, which is
shifted from original Te-fcc sub-lattice by (-$a$/4, -$b$/4, -$c$/4),
the anti-fluorite structure is obtained. Without distortions, the
Ag(1) and Ag(2) sublattices are equivalent, and the space group is
Fm3m with the inversion symmetry included. From the other point of
view, if we divide the large cube defined by $a, b, c$ into 8 small
cubes, only 4 out of 8 cube's body centers are occupied in zinc
blende, but they are now all occupied in anti-fluorite. All Ag atoms
in $\alpha$-Ag$_2$Te are tetrahedrally coordinated by four nearest
neighboring Te atoms.

\begin{figure}[tbp]
\includegraphics[clip,scale=0.3]{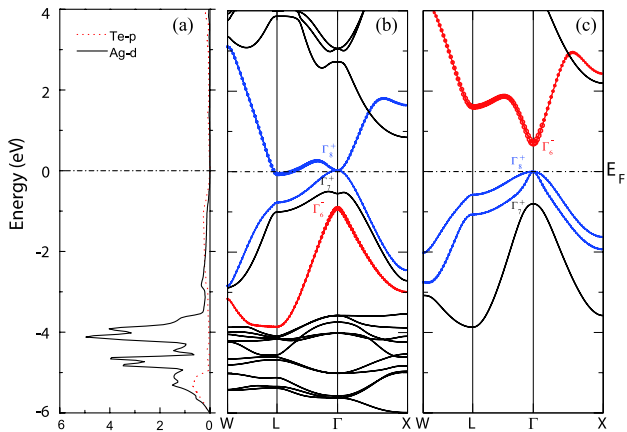}
\caption{(Color online) The calculated electronic structure of
  $\alpha$-Ag$_2$Te. (a) The projected density of states. (b) The
  original band structure and (c) the one after pushing Ag-$4d$ states
  artificially down to -20 eV. The projected component of Ag
  $s$-orbital is indicated as fat-bands.}
\label{band-alpha}
\end{figure}

Fig. 2 shows the calculated band structure and density of states of
$\alpha$-Ag$_2$Te (with optimized lattice parameter $a$=6.8\AA). The
electronic structure can be well understood as zero-gap semiconductor,
similar to HgTe, with inverted band ordering around the $\Gamma$
point. In conventional zinc blende semiconductor, the anion-$p$ states
(Te-$5p$) at $\Gamma$ point split into $\Gamma_8$ and $\Gamma_7$
manifolds due to the spin orbital coupling (SOC) with $\Gamma_8$
forming the valence band maximum (VBM). The conduction band minimum
(CBM) is mostly from the cation-$s$ state, called as $\Gamma_6$. The
$\Gamma_6$ is typically higher than the $\Gamma_8$ and $\Gamma_7$,
such as in CdTe, and a positive gap is formed. For HgTe, however, the
situation is different due to the presence of Hg-$5d$ states, which
are very shallow and hybridize with Te-$5p$ states strongly. Such
hybridization will push the Te-$5p$ upwards, resulting in an inverted
band structure with the $\Gamma_6$ state lower than the $\Gamma_8$
(therefore a negative band gap). Due to the double degeneracy of
$\Gamma_8$ manifolds, the zero gap semiconductor is formed. The
anti-fluorite structure of $\alpha$-Ag$_2$Te is similar to
zinc-blende, and its band structure is also very similar to HgTe. The
Te sites form the same fcc sub-lattice with similar lattice parameters
($a$=6.8 \AA~for Ag$_2$Te and $a$=6.46 \AA~for HgTe), but with two Ag
atoms instead of single Hg atom in the unit cell. The low energy
states of $\alpha$-Ag$_2$Te can be also characterized as $\Gamma_6,
\Gamma_7, \Gamma_8$. The Ag-$4d$ level is again very shallow, located
mostly from -6.0eV to -3.5eV as shown in Fig. 2. Since the Ag-$4d$
orbitals are less extended than the Hg-$5d$ orbitals, less $p$-$d$
hybridization may be expected. However, because there are two Ag sites
(instead of one Hg atom in HgTe) in one unit cell, the $p$-$d$
hybridizations are strong enough to push up the Te-$5p$ states and
leads to the inverted band structure (as shown in Fig. 2(b)) with
$\Gamma_6$ lower than $\Gamma_8$. To further demonstrate this
mechanism, we have shown in Fig. 2(c) the calculated band structure of
$\alpha$-Ag$_2$Te by artificially pushing the Ag-$4d$ states down to
-20 eV (out of the figure). Due to the reduced $p$-$d$ hybridization,
the Te-$4p$ states are now lower than the Ag-$5s$ states, giving a
positive band gap.

The above calculations were done based on the GGA, which is know to
underestimate the band gap of semiconductors (charge-transfer
gap). The situation now is different with Bi$_2$Te$_3$ and
Bi$_2$Se$_3$, where the gap is formed within the $p$ manifolds, and
mostly due to the SOC, which is a local physics and can be well
described by the GGA (or LDA)~\cite{Zhang}. The band gap problem for
HgTe and HgSe has been carefully studied by GW calculations and
semi-empirical method~\cite{HgS}, as well as comparing with
experiments.  It has been quantitatively suggested that the LDA
underestimate the band gap of HgTe (or HgSe) by the magnitude around
0.3$\sim$0.6eV~\cite{HgS}. Considering the strong similarity between
$\alpha$-Ag$_2$Te and HgTe as discussed above, the same size of error
bar from LDA (or GGA) may be expected. Nevertheless, even if the 0.6
eV correction for the band gap is added into our calculations, the
resulting band structure still supports the inverted band ordering at
$\Gamma$. This is because the calculated $\Gamma_6$ in GGA is far
below the $\Gamma_8$ (about -1.0 eV), a number much bigger than the
possible error bar for those compounds. Our results therefore suggest
that Ag$_2$Te has inverted band structure. This result is further
confirmed by our calculations using hybrid functional~\cite{HSE06}. As
long as the inverted band ordering remains for Ag$_2$Te, the
topological nature can be expected as discussed for
HgTe~\cite{Bernevig}.

To turn a cubic zero-gap compound into a true semiconductor with
finite band gap, certain distortion or strain has to be introduced to
break the symmetry and the degeneracy of $\Gamma_8$ manifolds. This
strategy has been followed for some of predicted topological
insulators, such as Heusler alloys~\cite{Heusler-1}, although
artificial distortions are not always easy. In our case, however, the
Ag$_2$Te undergoes structural distortion in its natural way: the high
temperature $\alpha$ phase changes into the $\beta$ phase below 417 K,
and the crystal structure of $\beta$-Ag$_2$Te can be understood as the
distorted anti-fluorite structure as shown in Fig. 1. Starting from
the cubic $\alpha$ phase with translational vectors $a, b, c$, we can
define three new translational vectors $A=a+b$, $B=-a/2-b/2+c$,
$C=a/2-b/2$. Then the distortion happens in such a way that both the
lengths of $A, B, C$ vectors and the angle between the $A$ and $B$
axis are varied, while keeping the $C$-axis perpendicular to the
$AB$-plane. The atomic positions are also shifted from their
high-symmetrical position (of cubic phase), resulting in the
monoclinic structure with space group $P2_1/c$~\cite{van}. Although
the distortions are a bit complicated, the main effect of distortions
in the electronic structure is to open up a gap around the Fermi
level. Fig. 3 shows the calculated band structure of $\beta$-Ag$_2$Te
with SOC included (using experimental structure~\cite{van}). It is
seen that the gap is around 80 meV, in good agreement with
experimental data~\cite{gap}. The present results can be also well
compared to earlier calculations~\cite{Ag2Te}, except the fact that
those studies neglected the SOC and therefore got metallic state.

\begin{figure}[tbp]
\includegraphics[clip,scale=0.4]{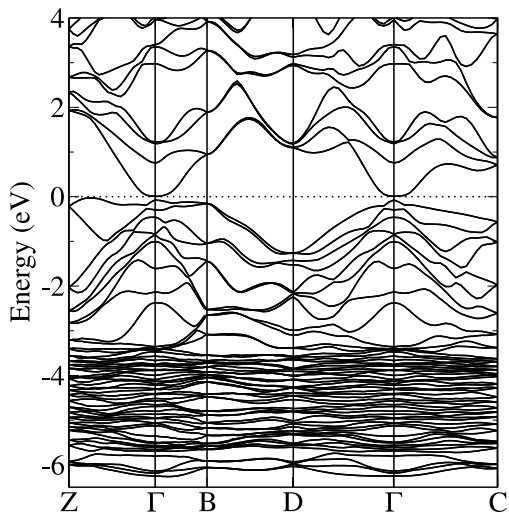}
\caption{(Color online) The calculated electronic structure of
  $\beta$-Ag$_2$Te including spin-orbit coupling. A gap is opened around
  the Fermi level (indicated by the horizontal dotted line).}
\label{band-beta}
\end{figure}

\begin{figure}[tbp]
\includegraphics[clip,scale=0.45]{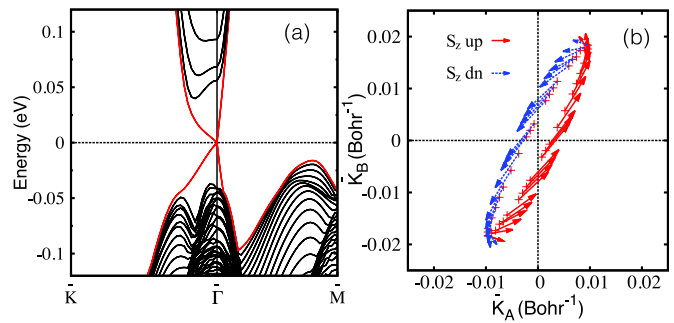}
\caption{(Color online) The surface states of $\beta$-Ag$_2$Te for the
  surface perpendicular to C-axis. (a) The surface band structure and
  Dirac cone calculated from a slab of 45 layers; (b) The Fermi
  surface and spin texture of surface states with chemical potential
  located 10 meV below Dirac point. The in-plane compents of spin are
  indicated as arrows, while the red (blue) color means the out of
  plane components pointing out-(in-) ward of the plane.}
\label{band-surf}
\end{figure}

The $\beta$-Ag$_2$Te with distorted structure is now true insulating,
but more importantly its topological nature is non-trivial due to the
inverted band structure. Since the $\beta$ phase has inversion
symmetry, we can identify its topological nature by analyzing the
parity of wave functions~\cite{Fu2}. We have calculated the parities
of occupied wave functions for time-reversal-invariant points in the
BZ, it is confirmed that the product of parities of occupied bands is
negative at $\Gamma$ point and positive for other points, which leads
to a $Z_2$=1 topological insulator~\cite{Fu2}.  This topological
nature should support gapless Dirac type surface states.  The
calculated surface state for the plane perpendicular to the $C$-axis
is shown in the Fig. 4.  It is clearly seen that we have single Dirac
cone on the surface similar to Bi$_2$Te$_3$, and
Bi$_2$Se$_3$~\cite{Zhang,Wei}. However, what is different is that the
surface Dirac cone is highly anisotropic, and the Fermi velocity
varies by about an order of magnitude moving around the cone. The
broken rotational symmetry in this case is due to the absence of
4-fold (in HgTe) or 3-fold (in Bi$_2$Se$_3$) rotational symmetry. The
spin direction is locked with lattice momentum and the spin chiral
texture shown in Fig. 4 (b) corresponds to the chemical potential
about 10 meV below the Dirac point. The spin direction of the surface
states has out-of-plane component, which gives us more freedom to
manipulate electron spin.

An effective $k\cdot p$ model can be established by considering only
the preserved time reversal symmetry for the surface state. The
possible model Hamiltonian around $\Gamma$ point is,
\begin{equation}
  H(k)=A\sigma_{x}+B\sigma_{y}+C\sigma_{z}+DI_{2\times2}.
\end{equation}
where $\sigma_{x}$, $\sigma_{y}$, $\sigma_{z}$ are Pauli matrix and
I$_{2\times2}$ is $2\times2$ identity
matrix. $A(k_{x},k_{y}),B(k_{x},k_{y}),C(k_{x},k_{y})$, and
$D(k_{x},k_{y})$ are functions of lattice momentum. Here it is noticed
that we have defined the principal axis $z$ to be along the
$C$-axis. In order to conserve the time reversal symmetry, $A$, $B$
and $C$ should contain only the odd terms of $k_{x}$ and $k_{y}$, and
$D$ contains the even terms. We expand $A$ and $B$ up to the third
order, $D$ up to the second order, while include only the linear term
for $C$ because the out-of-plane component of spin is much smaller
than the in-plane component,
\begin{equation}
  A(\vec{k})=c_{1}k_{x}+c_{2}k_{y}+c_{3}k_{x}^{3}+c_{4}k_{x}^{2}k_{y}+c_{5}k_{x}k_{y}^{2}+c_{6}k_{y}^{3}
\end{equation}
\begin{equation}
  B(\vec{k})=c_{7}k_{x}+c_{8}k_{y}+c_{9}k_{x}^{3}+c_{10}k_{x}^{2}k_{y}+c_{11}k_{x}k_{y}^{2}+c_{12}k_{y}^{3}
\end{equation}
\begin{equation}
  C(\vec{k})=c_{13}k_{x}+c_{14}k_{y}
\end{equation}
\begin{equation}
  D(\vec{k})=c_{15}k_{x}^{2}+c_{16}k_{y}^{2}+c_{17}k_{x}k_{y}.
\end{equation}

The parameters can be obtained by fitting the surface states
calculated from $ab-initio$. Both the shape and spin orientation of
surface states can be well reproduced with the parameters listed in
Ref.\cite{Para}.

The existence of gapless Dirac surface states in $\beta$-Ag$_2$Te
suggests that the observed unusual MR may have large contribution
coming from the surface or interface. Due to the large Landau level
spacing and high mobility, it can be estimated for $\beta$-Ag$_2$Te
that the quantum limit can be reached at about 10 K under only 10 Oe
field, in good agreement with experiment~\cite{Xu}. Our scenario is
further supported by the fact that experimental samples, doped with
excess Ag, are granular materials~\cite{Xu,Parish}, which makes the
surface/interface contribution significant. On the other hand, the
highly anisotropic surface states may cause large fluctuation of
mobility, which may also help to explain the unusual MR
behavior~\cite{Parish}. We have done similar calculations for
Ag$_2$Se, and found that it is also topologically non-trivial with
inverted band structure, while its $\beta$ phase crystal structure is
different with Ag$_2$Te.

In summary, we have shown by first principles calculations that
$\beta$-Ag$_2$Te is a new binary topological insulator provided by
nature, with highly anisotropic single surface Dirac cone. We suggest
that the observed unusual MR behavior can be understood from its
topological nature.  We acknowledge the supports from NSF of China and
that from the 973 program of China (No.2007CB925000).


\begin{references}
\bibitem{Para} The fitted parameters (with energy unit in eV and
  length unit in \AA) are: $c_1$=-1.192, $c_2$=0.781, $c_3$=-14.935,
  $c_4$=17.443, $c_5$=9.963, $c_6$=11.609, $c_7$=-1.807, $c_8$=0.822,
  $c_9$=-12.415, $c_{10}$=13.148, $c_{11}$=-55.148, $c_{12}$=19.788,
  $c_{13}$=-0.217, $c_{14}$=0.116, $c_{15}$=-1.559, $c_{16}$=2.687,
  $c_{17}$=-9.705.
\end{references}
\end{document}